\documentclass[%
 aip,
 jmp,%
 amsmath,amssymb,
 reprint,%
]{revtex4-2}

\usepackage{graphicx}
\usepackage{dcolumn}
\usepackage{bm}
\usepackage{mathptmx} 
\usepackage{amsmath,amssymb}
\begin{document}

\title[Localized Surface Plasmon Resonances of Simple Tunable Plasmonic Nanostructures]
{Localized Surface Plasmon Resonances of Simple Tunable Plasmonic Nanostructures} 

\author{Luke C. Ugwuoke}
\affiliation{Department of Physics, University of Pretoria.\\
Private bag X20, Hatfield 0028, South Africa}
\email{tjaart.kruger@up.ac.za}
\author{Tom\'{a}\v{s} Man\v{c}al}
\affiliation{Faculty of Mathematics and Physics, Charles University.\\
Ka Karlovu 5, 121 16 Prague 2, Czech Republic}
\author{Tjaart P. J. Kr\"{u}ger}
\affiliation{Department of Physics, University of Pretoria.\\
Private bag X20, Hatfield 0028, South Africa}
\email{tjaart.kruger@up.ac.za}

\date{\today}

\begin{abstract}
We derive and present systematic relationships between the analytical
formulas for calculation of the localized surface plasmon resonances
(LSPR) of some plasmonic nanostructures which we have categorized as
simple. These relationships, including some new formulas, are summarized
in a tree diagram which highlights the core-shell plasmons as the
generators of solid and cavity plasmons. In addition, we show that the LSPR of complex 
structures can be reduced to that of simpler ones, using the LSPR of a nanorice as a 
case study, in the dipole limit. 
All the formulas were derived using a combination of the Drude model, the
Rayleigh approximation, and the Fr\"{o}hlich condition. The formulas are
handy and they are in good agreement with the results of the plasmon
hybridization theory. The formulas also account for dielectric effects,
which provide versatility in the tuning of the LSPR of the nanostructures.
A simplified model of plasmon hybridization is presented, allowing us to
investigate the weak-coupling regimes of solid and cavity plasmons in the
core-shell nanostructures we have studied.
\end{abstract}

                             
\maketitle

\section{Introduction}\label{s1}
Single particle plasmonics \cite{Duyne13,Sam07,Sat16,Wu06,Norton16,Vince11,Kren12,Prodan04,Borgh09} 
is a subject of intense research due to its role in elucidating the 
intricate plasmonic behaviour of metallic nanostructures. In contrast to bulk studies, where particle heterogeneities and 
ensemble effects, such as random orientation of particles \cite{Duyne13}, limit the observation of specific plasmonic behaviour, 
single particle studies have been successful in revealing a number of interesting plasmonic properties of metallic 
nanostructures. 
The single particle studies identified  
polarization-dependent electric field enhancement in nanorods \cite{Sam07,Sat16} and nanoeggs \cite{Wu06,Norton16}, 
mode-mixing in nanoeggs \cite{Norton16}, and plasmon hybridization in nanoshells \cite{Prodan04}. 
Moreover, experimental studies at the single particle level 
have enabled the observation of multipolar modes, the so-called \emph{dark modes} \cite{Vince11,Kren12}, 
in nanodisks \cite{Kren12}, and material-dependent symmetric and antisymmetric plasmonic modes in nanorings \cite{Borgh09}.

The study of single particles in the Rayleigh regime, where particle sizes are 
small compared to the excitation wavelength,
has shown that the dipolar mode 
constitutes the largest contribution to the photoabsorption cross-section \cite{Duyne13,Sam07,Sat16}. 
For this reason, the dipolar mode is also known as the \emph{bright mode} \cite{Norton16,Vince11}. 
The particle shape exerts a harmonic force on the conduction electron gas in the metal, 
causing the collective electronic motion to exhibit 
resonant behaviour upon excitation near the surface plasmon frequency of the particle \cite{Sam07}. 
In addition, the electron cloud undergoes an intense polarization at the interface 
between the metal nanoparticle and the dielectric \cite{Duyne13,Sam07}. 
This resonant excitation of the collective oscillation of conduction electrons in metallic nanostructures is known as 
surface plasmon \cite{Duyne13,Sam07,Sat16}. 
Some nanostructures, such as spheres and spheroids, support particle-confined oscillations of conduction electrons,
which are referred to as localized surface plasmons (LSP) \cite{Sam07,Sat16}. On the other hand, planar nanostructures support 
propagating surface plasmons \cite{Sam07,Vince11}.
Drude metals such as noble metals, support surface plasmons due to their quasi-free electron-like behaviour. 
However, gold and silver nanostructures dominate plasmonics literature because their complex dielectric response 
lies in the visible and near-infrared regions of the electromagnetic spectrum, and 
were among the first to be reported experimentally via ellipsometry measurements \cite{Sam07,Sat16}.

The localized surface plasmon resonance (LSPR) of the dipolor mode usually manifests itself as the position of absorption maximum
in the photoabsorption cross-section of single particles \cite{Duyne13,Fend04}. 
It can reveal a lot of detail about the shape and size dependence of plasmonic behaviour in such particles. 
This observation has triggered both experimental and theoretical studies of
a slew of particle shapes of various sizes.
These shapes can be classified as solid or hollow metallic nanostructures with either smooth corners or sharp edges or even both.
The most common solid and hollow nanostructures with smooth corners include spheres and core-shell 
spheres \cite{Sam07,Sat16,Prodan04,Barnes16}, spheroids and core-shell 
spheroids \cite{Kren12,Barnes16,Moroz09,Birn90,Faraf11,Faraf96,Wang06},
spherical and spheroidal cavities \cite{Sam07,Johan08}, ellipsoids \cite{Sam07,Dalar17}, nanorods \cite{Prodan10,Vernon10}, 
and nanorings \cite{Borgh09}. Others include the reduced-symmetry core-shell nanostructures 
such as nanoeggs \cite{Norton16,Prodan04,Halas08}, nanocups \cite{Halas08}, and 
nanocaps \cite{Cortie05}. The most common solid nanostructures with sharp edges include 
nanocubes \cite{Xia05}, nanoprisms \cite{Van06}, nanostars \cite{Hafner06}, and nanoflowers \cite{Xu13}. However, crescent-shaped
nanostructures \cite{Vernon10} support both sharp edges and smooth corners. 
The most predominant and unique plasmonic behaviour in these structures include the ability to make some multipolar modes 
dipole-active via mode-mixing \cite{Norton16,Prodan04,Halas08,Hafner06} leading to multiple resonances, 
plasmon hybridization between solid and cavity plasmons \cite{Prodan04,Borgh09,Wang06}, 
incident electric field polarization-dependent LSPR \cite{Wang06,Prodan10,Vernon10}, enhanced local fields comparable to those in 
nanosphere dimers \cite{Wu06,Norton16,Xia05,Van06,Hafner06,Xu13}, red-shifted absorption spectra \cite{Cortie05,Hafner06},
and the ability to suppress plasmon damping \cite{Duyne13}.

The LSPR of the afore-mentioned nanostructures can be tuned via size, dielectric embedding medium, and material composition, 
in order to achieve the desired effect in applications where an optimal LSPR is required. Applications where such tunable
nanostructures are needed include 
particle tracking via fluorescence nanoscopy \cite{Duyne13}, imaging of live cells via surface-enhanced Raman 
spectroscopy \cite{Xu13}, biosensing and chemical sensing via LSPR spectroscopy \cite{Fend04,Haf06,Yury19}, 
enhanced photoabsorption and free-carrier generation in solar cells \cite{Polman08,Yang12,Lee13}, 
metal-enhanced fluorescence in hybrid nanostructures \cite{Vince11,Sayed00,Yury19,Novot07,Martin12}, coherent propagation of 
electromagnetic energy in plasmon waveguides \cite{Maier01}, and a proposed localized and targeted photothermal therapy for cancer 
treatment \cite{Drez11}.

\begin{figure}[h!]
\centering
\includegraphics[width = .7\linewidth]{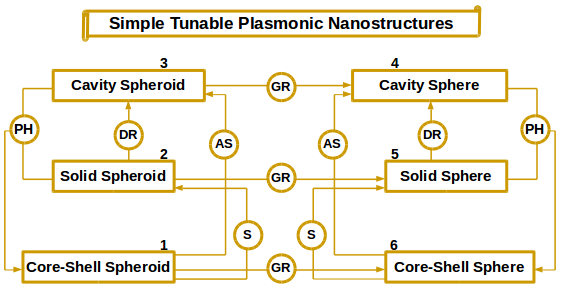}
\caption{\small{A tree diagram illustrating the relationships between the LSPR of the nanostructures we have studied. 
\textbf{PH} stands for Plasmon Hybridization, \textbf{DR} for Dielectric Reversal, \textbf{GR} for Geometric Reduction, 
\textbf{S} for Symmetrization, and \textbf{AS} for Anti-Symmetrization.} }
\label{f1}
\end{figure}

Despite their differences in shape, single metallic nanostructures of different shapes have been shown to display some 
common trends in their plasmonic behaviour. For example, plasmon hybridization of cavity and solid plasmons, and the
presence of symmetric and antisymmetric modes in nanoshells \cite{Prodan04}, nanorings \cite{Borgh09}, and 
nanorice \cite{Wang06}. In this report, we took this further by exploring other possible relationships between 
the members of a group of tunable plasmonic nanostructures which we have categorized as \emph{simple}.
This classification is motivated by three typical characteristics: the nanostructures have smooth corners, 
their dipolar LSPR can be expressed analytically via some handy formulas,  
and some relationships exist between them according to FIG. \ref{f1}. 

\begin{figure}
\centering
\includegraphics[width = 0.50\linewidth]{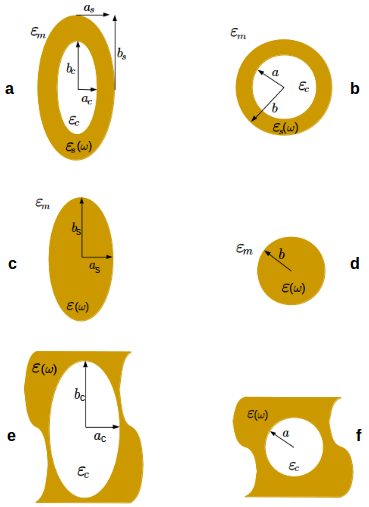}
\caption{\small{Model geometries of the nanostructures we have classified as \emph{simple}. 
(\textbf{a}) A core-shell prolate spheroid of aspect ratio
$q_{s} = b_{s}/a_{s}$, with a metallic shell of non-uniform thickness and dielectric 
material $\varepsilon_{s}(\omega)$, and a confocal core of aspect ratio $q_{c} = b_{c}/a_{c}$. 
(\textbf{b}) A core-shell sphere of aspect ratio $a/b$, with a metallic shell of uniform thickness and dielectric 
material $\varepsilon_{s}(\omega)$, and a concentric core.  
Each core-shell nanostructure is embedded in a homogeneous dielectric medium of dielectric constant $\varepsilon_{m}$, and 
the core is filled with a dielectric of dielectric constant $\varepsilon_{c}$. 
(\textbf{c}) A solid prolate spheroid of aspect ratio $b_{s}/a_{s}$. 
(\textbf{d}) An isotropic solid sphere of radius $b$. 
Each solid nanostructure consists of a metal of dielectric material $\varepsilon(\omega)$, 
embedded in a homogeneous dielectric medium of dielectric constant $\varepsilon_{m}$. 
(\textbf{e}) A prolate spheroidal cavity of aspect ratio $b_{c}/a_{c}$. 
(\textbf{f}) An isotropic spherical cavity of radius $a$. 
Each cavity is formed in a homogeneous infinite metallic medium of dielectric material $\varepsilon(\omega)$, and
filled with a dielectric of dielectric constant $\varepsilon_{c}$. 
The surface charge distributions on the fundamental nanostructures, and
in the symmetric and antisymmetric modes of the core-shell nanostructures, 
in the presence of a uniform electric field polarized along the x-axis, are available in the ESI (see S 3). 
They were drawn following the convention of Ref. \cite{Prodan04}.
}}\label{f2}
\end{figure}

The main objective of this study is to show that certain relationships exist between the LSPR of 
the members of a group of simple plasmonic nanostructures. In addition, 
the multipolar and dipolar LSPR of a nanorice (a core-shell prolate spheroid) will be derived. 
Likewise, we will show that the dipolar LSPR of a nanorice can be used to derive the dipolar LSPR of a 
nanoshell (a core-shell sphere) and the dipolar LSPR of other nanostructures in FIG. \ref{f1}. We shall make use of
the relationships depicted in FIG. \ref{f1}. These derivations will include dielectric effects. 
Finally, we shall present a simplified model of plasmon hybridization in the core-shell nanostructures 
in FIG. \ref{f1} and FIG. \ref{f2}.

\section{Theoretical Methods}\label{s2}
Several methods have been proposed for the derivation of the LSPR of metallic nanostructures.
These methods rely on various approximation schemes depending on the complexity of the target
structure. Nevertheless, the results reported for a given structure via these different methods
have been very consistent. 
Pioneer schemes focus mostly on nanostructures with smooth corners, where analytical solutions are possible.
Examples of such schemes include the local density approximation \cite{Lund82,Land03}, 
the time-dependent local density approximation \cite{Prodan10}, and the plasmon hybridization theory \cite{Prodan04}.  
Others include an analytical solution of the Laplace equation in order to obtain the static 
polarizability \cite{Sam07,Sat16,BoHu08},
analytical solution of Maxwell's equations in order to obtain the dispersion relation for LSP waves \cite{Anasta16} or to obtain
the dynamic polarizability of the nanostructure (Mie theory) \cite{Raza14,BoHu08,Dmitri13}. 
On the other hand, the post-pioneer schemes, which focus mostly on nanostructures 
with arbitrary shapes, rely completely on numerical methods. These schemes use numerical methods to solve the 
Laplace equation or Maxwell's equations, in order to obtain the static or dynamic polarizability of the 
nanostructure, respectively. 
Examples of such numerical schemes include the electrostatic eigenmode method \cite{Vernon10}, 
the boundary element method \cite{Abajo08}, the discrete dipole approximation method \cite{Van06,Abajo08}, and 
the finite difference time domain method \cite{Halas08,Xia05,Abajo08}. 
Schemes based on transformation optics have also been reported \cite{Zayats13}. 
Once the polarizability of the target structure is known, the extinction cross-section can be 
calculated \cite{BoHu08,Dmitri13}.  
Most of the above schemes predict the LSPR of the target nanostructures from their extinction cross-section 
or from their absorption cross-section for small particles. 
We shall use one of the pioneer schemes: a combination of the Drude model \cite{Sam07}, 
the Rayleigh approximation, also called the quasistatic approximation \cite{Birn90,Faraf11}, 
and the LSP resonance condition, also called the Fr\"{o}hlich condition \cite{Sam07,Barnes16,Dalar17}, 
to derive handy formulas for the prediction of the LSPR of some simple nanostructures.

In the local response approximation (LRA) \cite{Sam07,Barnes16,Raza14}, the dielectric response of most metals can be 
described via a complex and wavevector-independent dielectric function given by the the Drude model: 
\begin{equation}\label{e1}
\varepsilon(\omega) = \varepsilon_{\infty} - \frac{\omega^{2}_{p}}{\omega(\omega+i\gamma)}.
\end{equation}
Eq. (\ref{e1}) accounts for the polarization of the positive ion core in the metal via the metal background dielectric response, 
$\varepsilon_{\infty}$, which depends weakly on frequency, 
as well as intra-band effects due to the polarization of the free electron plasma. 
The quantity $\omega_{p}$ is the bulk plasma frequency of the free electrons in the metal, $\gamma$ is the damping rate of the free electrons, and $\omega$ is the 
frequency of the incident radiation. In the derivation of the LSPR, only the real part of Eq. (\ref{e1}) 
is retained \cite{Sam07,Barnes16}, since the LSPR is a measurable physical quantity.
Upon resonant excitation, i.e., $\omega \approx \omega_{r}$, one obtains the resonance frequency $\omega_{r}$ from the real part of 
Eq. (\ref{e1}) as: 
\begin{equation}\label{e2}
\omega_{r} = \sqrt{\frac{\omega^{2}_{p}}{\varepsilon_{\infty}-\Re[\varepsilon(\omega_{r})]}-\gamma^{2}},
\end{equation}
Following the approach of Wu et al. \cite{Wu06}, we shall ignore the free-electron damping term in 
Eq. (\ref{e2}) (So that $\Re[\varepsilon(\omega_{r})] = \varepsilon(\omega_{r})$). 
Their approach suggests that using a renormalized value of the bulk plasma frequency of gold, 
without considering free-electron damping, reproduces observed values of the LSPR of nanoshells over a wide range of 
aspect ratios. 
In this work, we have chosen to consider gold nanostructures only with the values $\omega_{p} = 4.6$~eV, 
which is the renormalized bulk plasma frequency of gold given in Ref. \cite{Wu06}, and $\varepsilon_{\infty} = 1$.

At resonance, the expression for $\Re[\varepsilon(\omega_{r})]$ in Eq. (\ref{e2}) 
differs in different nanostructures making it possible for 
the LSPR to be tuned. For a given nanostructure, $\Re[\varepsilon(\omega_{r})]$ is obtained by imposing the resonance condition 
on its multipole polarizability. The electrostatic polarizability is a complex quantity that determines the ease at which the 
surface charges on the nanostructure are polarized upon the application of an electric field. It depends on size, shape, 
material composition, and dielectric environment of the nanostructure, as well as on the frequency of the incident radiation. 
It also depends on the polarization of the incident electric field in some nanostructures such as nanodisks \cite{Kren12}, 
nanorods \cite{Prodan10}, nanorice \cite{Wang06}, and nanoeggs \cite{Wu06}.
It is an important physical quantity in plasmonics, because the photoabsorption cross-section is proportional to the imaginary 
part of the polarizability \cite{Sam07,Land03,Raza14}.

Given some multipole polarizability $\alpha_{lm}(\omega)$, the 
Fr\"{o}hlich condition requires that the real part of the denominator of $\alpha_{lm}(\omega)$ vanishes 
at resonance \cite{Sam07,Barnes16,Dalar17,Neeves89}. 
Here, $l$ is an angular momentum number that indicates the number of multipoles, and $m$ is an azimuthal number that 
indicates either a parallel or perpendicular polarization of the incident field. All nanostructures in this work have 
azimuthal symmetry, in which case the $m$ dependence is ignored.

By obtaining the amplitude of the electrostatic 
potential induced on the surface of a given nanostructure (also referred to as the amplitude of the 
\emph{scattered potential} \cite{Faraf96}) in terms of the amplitude of the external potential, one obtains the complex 
multipole polarizability of the nanostructure. Its denominator is therefore used to solve for $\Re[\varepsilon(\omega_{r})]$.

Detailed derivations of all the formulas provided below are given in the electronic supplementary information (ESI). 
The symbol ``$|~|$'' stands for determinant wherever it appears. For convenience, the subscript ``$r$'' in Eq. (\ref{e2}) 
has been dropped.

\section{Results and Discussion}\label{s3}
\subsection{LSPR of Simple Nanostructures}\label{s3.1}
In this section, we first derive the LSPR of metallic shell-dielectric core nanostructures  
with the inclusion of the dielectric constants $\varepsilon_{\infty}, \varepsilon_{c}$ and $\varepsilon_{m}$. 
The nanostructures considered include a core-shell spheroid and a core-shell sphere. 
The plasmonic modes of these nanostructures can be referred to as the \emph{hybrid plasmon modes}, 
since they are formed from hybridization of the fundamental plasmon modes \cite{Prodan04,Wang06,Wu06}.
For brevity, the LSPR of a core-shell sphere was obtained from that of a core-shell spheroid, 
using Geometric Reduction (\textbf{GR}), in the dipole limit only.

Secondly, we derived the LSPR of a solid spheroid, a cavity spheroid, a solid sphere, and a cavity sphere, 
from the LSPR of the core-shell nanostructures, using the relationships in FIG. \ref{f1}, 
except for Dielectric Reversal (\textbf{DR}), 
which was performed using some results derived for the solid plasmons. 
Detailed derivations of their LSPR using the approach discussed in Section \ref{s2},
which correspond to the formulas we reported here, are available in the ESI. 
Solid prolate spheroids are approximations of nanorods \cite{Wang06}, while a solid oblate spheroid is also 
known as a nanodisk \cite{Kren12} or platelet \cite{Barnes16}. The plasmonic modes of these nanostructures can be 
referred to as the \emph{fundamental plasmon modes}, due to the role they play in plasmon hybridization \cite{Prodan04,Wang06}.

To obtain the LSPR of these fundamental nanostructures from those of the corresponding core-shell structures, 
we made use of the fact that the symmetric mode of a core-shell nanostructure is formed from symmetric coupling of 
solid and cavity plasmons, while the antisymmetric mode is formed from antisymmetric coupling of the 
solid and cavity plasmons \cite{Wu06,Prodan04,Wang06}. However, a symmetric de-coupling of the symmetric and anti-symmetric 
modes in the core-shell nanostructure leads to the LSPR of the solid plasmons, while an antisymmetric de-coupling 
of the symmetric and antisymmetric modes leads to the LSPR of the cavity plasmons. The former process is what we have chosen 
to refer to as Symmetrization (\textbf{S}), and the latter as Anti-Symmetrization (\textbf{AS}).

The results we obtained for the fundamental nanostructures are already well-known but we present them to show that they 
can also be obtained using the relationships depicted in FIG. \ref{f1}. 

\subsubsection{Core-Shell Spheroid}\label{s3.1.1} 
Using the solutions of the Laplace equation in the medium, shell, and core regions of a core-shell spheroid as given in
Refs. \cite{Birn90,Faraf11,Scaife99}, for the longitudinal polarization, 
we obtained the longitudinal multipole polarizability of the core-shell 
spheroid (see (S.52)-(S.65), ESI), and applied the approach described in Section \ref{s2}, to obtain the longitudinal 
multipole LSPR of the core-shell prolate spheroid as:
\begin{equation}\label{e3}
\resizebox{.9\hsize}{!}{$
\omega^{\parallel}_{l\pm} = \frac{\omega_{p}}{\sqrt{2}}\sqrt{
\frac{\begin{vmatrix}
2\varepsilon_{\infty} & -1\\
[\beta+\zeta-\Omega_{1}-\Omega_{3}] & (\Omega-\Delta)
\end{vmatrix}
\pm \sqrt{ 
\begin{vmatrix}
\beta & 1\\
\zeta & 1
\end{vmatrix}^{2}
+ \begin{vmatrix}
(\Omega_{1}+\Omega_{2}) & -4\\
[\Lambda\Omega-\Omega_{2}(\Omega-\Delta)] & [\Omega_{1}+\Omega_{2}-2(\beta+\zeta)]
\end{vmatrix}
}
}
{\begin{vmatrix}
[\varepsilon_{\infty}^{2}(\Omega-\Delta)+(\Omega_{2}-\Lambda)] & -\varepsilon_{\infty}\\
[\beta+\zeta-\Omega_{1}-\Omega_{3}]                            & 1
\end{vmatrix}
}
}
$}.
\end{equation}
Here, $\omega^{\parallel}_{l-}$ and $\omega^{\parallel}_{l+}$ are the longitudinal symmetric and antisymmetric multipolar modes 
respectively, and 
\begin{equation}\label{e4}
\Omega \equiv Q_{l}(v_{s})Q^{\prime}_{l}(v_{s}),~
\Omega_{1} \equiv \varepsilon_{m}\Omega,~
\Omega_{2} \equiv \varepsilon_{c}\Omega_{1},~
\Omega_{3} \equiv \varepsilon_{c}\Omega, 
\end{equation}
\begin{equation}\label{e5}
\resizebox{1.\hsize}{!}{$
\Lambda \equiv \varepsilon_{m}\varepsilon_{c}\frac{Q^{\prime}_{l}(v_{s})Q_{l}(v_{c})P_{l}(v_{s}) }{P_{l}(v_{c})},~
\beta \equiv \varepsilon_{m}\frac{Q^{\prime}_{l}(v_{s})Q^{\prime}_{l}(v_{c})P_{l}(v_{s}) }{P^{\prime}_{l}(v_{c})},~
\zeta \equiv \varepsilon_{c}\frac{Q_{l}(v_{s})Q_{l}(v_{c})P^{\prime}_{l}(v_{s}) }{P_{l}(v_{c})},
\Delta \equiv \frac{Q_{l}(v_{s})Q^{\prime}_{l}(v_{c})P^{\prime}_{l}(v_{s}) }{P^{\prime}_{l}(v_{c})},
$}
\end{equation}
where $P_{l}(v)$ and $Q_{l}(v)$ are Legendre functions of the first and second kind, respectively, 
$v$ is the radial coordinate of the spheroid, the subscripts $c$ and $s$ denote core and shell, respectively, 
and ``$\prime$'' stands for derivative.
Eq. (\ref{e3}) can be transformed to the LSPR of a core-shell oblate spheroid by evaluating Eqs. (\ref{e4})  and (\ref{e5}) 
with $v = iv$ \cite{Birn90,Faraf11}. 
However, in the dipole limit, this transformation is done through the geometric factors \cite{Barnes16,BoHu08}.
In the dipole limit, we evaluated Eqs. (\ref{e4}) and (\ref{e5}) for $l = 1$, and substituted the results in Eq. (\ref{e3}) 
to obtain the longitudinal dipolar LSPR of the core-shell spheroid as:
\begin{equation}\label{e6}
\resizebox{1.\hsize}{!}{$
\omega^{\parallel}_{1\pm} = \frac{\omega_{p}}{\sqrt{2}}\sqrt{
\frac{\begin{vmatrix}
[2L_{s}^{\parallel}\varepsilon_{\infty}+\varepsilon_{m}(1-L_{c}^{\parallel}-L_{s}^{\parallel})] & 
[-L_{s}^{\parallel}[L_{c}^{\parallel}+f_{c}(1-L_{s}^{\parallel})]] \\
(\varepsilon_{m}+\varepsilon_{c}-2\varepsilon_{\infty}) & 1
\end{vmatrix}
\pm \sqrt{ 
\begin{vmatrix}
\varepsilon_{m}(1-L_{c}^{\parallel}) & \varepsilon_{c}L_{c}^{\parallel} \\
L_{s}^{\parallel}  &  1-L_{s}^{\parallel}
\end{vmatrix}^{2}
+ f_{c}L_{s}^{\parallel}(1-L_{s}^{\parallel})\begin{vmatrix}
                   L_{s}^{\parallel}[2L_{c}^{\parallel}+f_{c}(1-L_{s}^{\parallel})] & -2\varepsilon_{m} \\
                   [\varepsilon_{m}+\varepsilon_{c}+(L_{c}^{\parallel}+L_{s}^{\parallel})(\varepsilon_{c}-\varepsilon_{m})]
                   &  (\varepsilon_{c}-\varepsilon_{m})^{2}
                  \end{vmatrix}
}
}
{\begin{vmatrix}
[\varepsilon_{m}(1-L_{s}^{\parallel})+\varepsilon_{\infty}L_{s}^{\parallel}] & -f_{c}L_{s}^{\parallel}(1-L_{s}^{\parallel})\\
[\varepsilon_{\infty}(\varepsilon_{m}+\varepsilon_{c}-\varepsilon_{\infty})-\varepsilon_{m}\varepsilon_{c}]
& [\varepsilon_{\infty}(1-L_{c}^{\parallel})+\varepsilon_{c}L_{c}^{\parallel}]
\end{vmatrix}
}
}
$}.
\end{equation}
Here, we have
\begin{equation}\label{e7}
L^{\parallel} = [(v\coth^{-1}v) -1](v^{2}-1),~v = \frac{1}{\sqrt{1-q^{-2}}},~q > 1,~~
f_{c} = \frac{v_{c}(v^{2}_{c}-1)}{v_{s}(v^{2}_{s}-1)} ~~~~~\text{(Prolate)},
\end{equation}
\begin{equation}\label{e8}
L^{\parallel} = \frac{1}{2}[(v^{2}+1)(v\cot^{-1}v) -v^{2}],~v = \frac{1}{\sqrt{q^{-2}-1}},~0<q < 1,~~
f_{c} = \frac{v_{c}(v^{2}_{c}+1)}{v_{s}(v^{2}_{s}+1)} ~~~~~\text{(Oblate)}, 
\end{equation}
where $L^{\parallel}$ is the longitudinal geometric factor of the spheroid, $q$ is its aspect ratio, and 
$f_{c}$ is the core-volume fraction of the core-shell spheroid. The parameter $q$ is evaluated on the core and shell 
to obtain the values of $L^{\parallel}$ and $v$ on the core and shell, respectively, using
\begin{equation}\label{e9}
q_{c} = \frac{b_{c}}{a_{c}},~b_{c}>a_{c},~ q_{s} = \frac{b_{s}}{a_{s}},~b_{s}>a_{s},~q_{c}>q_{s} ~~~ \text{(Prolate)},
\end{equation}
\begin{equation}\label{e10}
q_{c} = \frac{b_{c}}{a_{c}},~b_{c}<a_{c},~ q_{s} = \frac{b_{s}}{a_{s}},~b_{s}<a_{s},~q_{c}<q_{s} ~~~ \text{(Oblate)}. 
\end{equation}
In the limit of $\varepsilon_{c} = \varepsilon_{m} = \varepsilon_{\infty} = 1$, Eq. (\ref{e6}) reduces to 
\begin{equation}\label{e11}
\resizebox{.6\hsize}{!}{$
\omega^{\parallel}_{1\pm} = \frac{\omega_{p}}{\sqrt{2}}\sqrt{
\Big(1-L_{c}^{\parallel}+L_{s}^{\parallel}\Big) \pm \Big(1-L_{c}^{\parallel}-L_{s}^{\parallel}\Big)
\sqrt{ 1 + \frac{ 4f_{c}L_{s}^{\parallel}\Big(1-L_{s}^{\parallel}\Big) }{ \Big(1-L_{c}^{\parallel}-L_{s}^{\parallel}\Big)^{2} }}  
}.
$}
\end{equation}

For transverse polarization, the transverse geometric factor of the spheroid, $L^{\perp}$, 
is obtained from the sum rule \cite{BoHu08,Moroz09} as follows:
\begin{equation}\label{e12}
L^{\perp} = \frac{1}{2}\Big(1-L^{\parallel}\Big).
\end{equation}
The transverse dipolar LSPR of the core-shell spheroid is obtained by replacing $L^{\parallel}$  with $L^{\perp}$ in 
Eqs. (\ref{e6}) and (\ref{e11}), respectively.
\subsubsection{Core-Shell Sphere}\label{s3.1.2}
Using the solutions of the Laplace equation in the medium, shell, and core regions of a core-shell sphere as given in
Refs. \cite{Sat16,Bott73,BoHu08,Neeves89}, we obtained the multipole polarizability of the core-shell sphere 
(see (S.35)-(S.48), ESI), and applied the approach described in Section \ref{s2}, 
to obtain the multipole LSPR of the core-shell sphere as:
\begin{equation}\label{e13}
\resizebox{1.\hsize}{!}{$
\omega_{l\pm} = \frac{\omega_{p}}{\sqrt{2}}\sqrt{
\frac{\begin{vmatrix}
l(l+1) & -[l^{2}\varepsilon_{c}+(l+1)^{2}\varepsilon_{m}]\\
1 & [2\varepsilon_{\infty}+q^{2l+1}(\varepsilon_{m}+\varepsilon_{c}-2\varepsilon_{\infty})]
\end{vmatrix}
\pm \sqrt{ 
\begin{vmatrix}
(l+1)^{2} & \varepsilon_{c}\\
l^{2} & \varepsilon_{m}
\end{vmatrix}^{2}
+ l(l+1)q^{2l+1}\begin{vmatrix}
[l(l+1)q^{2l+1}(\varepsilon_{c}-\varepsilon_{m})^{2}] & -2\\
[l\varepsilon_{c}+(l+1)\varepsilon_{m}]^{2}+\varepsilon_{c}\varepsilon_{m}(2l+1)^{2} & 1
\end{vmatrix}
}
}
{\begin{vmatrix}
[(l+1)\varepsilon_{m}+l\varepsilon_{\infty}] & -l(l+1)q^{2l+1}\\
[\varepsilon_{\infty}(\varepsilon_{m}+\varepsilon_{c}
-\varepsilon_{\infty})-\varepsilon_{m}\varepsilon_{c}] & [l\varepsilon_{c}+(l+1)\varepsilon_{\infty}]
\end{vmatrix}
}
},
$}
\end{equation}
where $q = a/b$ is the aspect ratio of the nanoshell, and $\omega_{l-}$ and $\omega_{l+}$ are the 
symmetric and antisymmetric multipolar modes, respectively. The $\varepsilon_{c} = \varepsilon_{m} = \varepsilon_{\infty} = 1$
limit of Eq. (\ref{e13}) is very popular in plasmonic literature \cite{Sam07,Sat16,Prodan04}.

As depicted in Fig. (\ref{f1}), we can obtain the dipolar LSPR of the core-shell sphere
i.e the dipole limit of Eq. (\ref{e13}), from the dipolar LSPR of a core-shell
spheroid via \textbf{GR}. To do this, we need the 
geometric factor of an isotropic sphere. It is polarization independent \cite{Barnes16,BoHu08} and is obtained from the 
sum rule giving 
\begin{equation}\label{e14}
L = 1/3. 
\end{equation}
Eq. (\ref{e14}) allows us to reduce the spheroid to a sphere. By replacing the geometric factors in Eq. (\ref{e6}) 
with Eq. (\ref{e14}), and redefining $f_{c}$ as $(a/b)^{3}$, where $f_{c}$ is the core-volume fraction of the 
nanoshell, one obtains the dipolar LSPR of the nanoshell as: 
\begin{equation}\label{e15}
\resizebox{.8\hsize}{!}{$
\omega_{1\pm} = \frac{\omega_{p}}{\sqrt{2}}\sqrt{
\frac{\begin{vmatrix}
3 & -(1+2f_{c}) \\
(\varepsilon_{m}+\varepsilon_{c}-2\varepsilon_{\infty}) & (\varepsilon_{m}+2\varepsilon_{\infty})
\end{vmatrix}
\pm \sqrt{ 
\begin{vmatrix}
4 & \varepsilon_{c}\\
1 & \varepsilon_{m}
\end{vmatrix}^{2}
+ 4f_{c}\begin{vmatrix}
                   (1+f_{c}) & -3\varepsilon_{m} \\
                   (\varepsilon_{m}+5\varepsilon_{c})  &  (\varepsilon_{c}-\varepsilon_{m})^{2}
                  \end{vmatrix}
}
}
{\begin{vmatrix}
(2\varepsilon_{m}+\varepsilon_{\infty}) & -2f_{c}\\
[\varepsilon_{\infty}(\varepsilon_{m}+\varepsilon_{c}-\varepsilon_{\infty})-\varepsilon_{m}\varepsilon_{c}]
& (\varepsilon_{c}+2\varepsilon_{\infty}) 
\end{vmatrix}
}
},
$}
\end{equation}
where $\omega^{-}_{1}$ and $\omega^{+}_{1}$ are the symmetric and antisymmetric dipolar modes, respectively. 

In the limit of $\varepsilon_{c} = \varepsilon_{m} = \varepsilon_{\infty} = 1$, Eq. (\ref{e15}) reduces to the  
well-known result: 
\begin{equation}\label{e16}
\omega^{\pm}_{1} = \frac{\omega_{p}}{\sqrt{2}}\sqrt{1 \pm \frac{1}{3}\sqrt{1+8f_{c} }}.
\end{equation}

\subsubsection{Solid Spheroid} 
In the core-shell spheroid discussed in Section \ref{s3.1.1}, the quantity $\Omega$, defined in 
Eq. (\ref{e4}), plays the role of the coupling constant between the cavity and solid plasmons of the core-shell spheroid in the multipole limit. 
This is because by setting $\Omega = 0$ in the symmetric mode $(\omega^{\parallel}_{l-})$ of Eq. (\ref{e3}), and 
making use of Eqs. (\ref{e4}) and (\ref{e5}), one obtains the longitudinal multipole LSPR of the solid prolate spheroid as: 
\begin{eqnarray}
\omega^{s\parallel}_{l} = \omega^{\parallel}_{l-}(\Omega = 0)~~~~~~~~~~~~~~~~~~~~~~~~~~~~~~~~~~~~~~~~~~~~~~~~~~~~~~ \nonumber \\
= \omega_{p}\sqrt{\frac{P^{\prime}_{l}(v_{s})Q_{l}(v_{s})}
{\varepsilon_{\infty}P^{\prime}_{l}(v_{s})Q_{l}(v_{s})-\varepsilon_{m}P_{l}(v_{s})Q^{\prime}_{l}(v_{s}) }}.
~~~~~~~~~~~\label{e17}
\end{eqnarray}
However, in the dipole limit, the core-volume fraction $f_{c}$ plays the role of the coupling constant. 
This is because by setting $f_{c} = 0$ in the symmetric dipolar mode $(\omega^{\parallel}_{1-})$ of Eq. (\ref{e6}), 
the longitudinal dipolar LSPR of the solid prolate spheroid is found: 
\begin{eqnarray}
\omega_{1}^{s\parallel} = \omega^{\parallel}_{1-}(f_{c} = 0)~~~~~~~~~~~~~~~~~~~~~~~~~~~~~~~~~~~~~~~~~~~ \nonumber \\
= \omega_{p}\sqrt{\frac{L^{\parallel}_{s}}{\varepsilon_{\infty}L^{\parallel}_{s}
+\varepsilon_{m}\Big(1-L^{\parallel}_{s}\Big) }}.~~~~~~~~~~~~~~~~~~~~~~~\label{e18}
\end{eqnarray}
This process corresponds to \textbf{S}. 
Eq. (\ref{e18}) is also the dipole limit of Eq. (\ref{e17}). 
Note that the $\varepsilon_{m} = \varepsilon_{\infty} = 1$ limit of Eq. (\ref{e18}) can also be obtained by performing \textbf{S} of Eq. (\ref{e11}).

A similar equation for the solid oblate spheroid can be obtained by making use of Eq. (\ref{e8}).
In the transverse polarization, the transverse dipolar LSPR of the solid prolate and oblate spheroids can be respectively 
obtained by making use of Eq. (\ref{e12}).

\subsubsection{Cavity Spheroid}
In this section we use \textbf{DR} to obtain the LSPR of a cavity prolate spheroid \cite{Sam07}.
To do this, the expression obtained for $\Re[\varepsilon(\omega_{r})]$ in the case of a solid prolate spheroid is used, but the positions of the dielectric constants are reversed. Also, $\varepsilon_{m}$ is replaced with 
$\varepsilon_{c}$, and $v_{s}$ with $v_{c}$ to obtain a new expression for 
$\Re[\varepsilon(\omega_{r})]$ (see SII.4, ESI), which is substituted in Eq. (\ref{e2}) to give the 
longitudinal multipole LSPR of the cavity prolate spheroid as: 
\begin{eqnarray}
\omega^{c\parallel}_{l} = \omega^{\parallel}_{l+}(\Omega = 0) ~~~~~~~~~~~~~~~~~~~~~~~~~~~~~~~~~~~~~~~~~~~~ \nonumber \\
~~~~ = \omega_{p}\sqrt{\frac{P_{l}(v_{c})Q^{\prime}_{l}(v_{c})}
{\varepsilon_{\infty}P_{l}(v_{c})Q^{\prime}_{l}(v_{c})-\varepsilon_{c}P^{\prime}_{l}(v_{c})Q_{l}(v_{c}) }}.~\label{e19}
\end{eqnarray}
Eq. (\ref{e19}) can also be obtained by performing \textbf{AS} of Eq. (\ref{e3}), 
i.e, by setting $\Omega = 0$ in the antisymmetric mode $(\omega_{l+})$ of Eq. (\ref{e3}).
The dipole limit of Eq. (\ref{e19}) can be obtained by performing \textbf{AS} of Eq. (\ref{e6}) to 
obtain:
\begin{eqnarray}
\omega_{1}^{c\parallel} = \omega^{\parallel}_{1+}(f_{c} = 0)~~~~~~~~~~~~~~~~~~~~ \nonumber \\
~~~~~~ = \omega_{p}\sqrt{\frac{1-L^{\parallel}_{c}}{\varepsilon_{\infty}\Big(1-L^{\parallel}_{c}\Big)
+\varepsilon_{c}L^{\parallel}_{c} }}.\label{e20}
\end{eqnarray}

A similar equation for the cavity oblate spheroid can be obtained by making use of Eq. (\ref{e8}). 
Note that the $\varepsilon_{c} = \varepsilon_{\infty} = 1$ limit of  
Eq. (\ref{e20}) can also be obtained by performing \textbf{AS} of Eq. (\ref{e11}).
The transverse LSPR of the cavity prolate and oblate spheroids can be obtained by making use of Eq. (\ref{e12}).

\subsubsection{Solid Sphere}
To obtain the multipole LSPR of the solid sphere from that of a core-shell sphere, we performed \textbf{S} 
on Eq. (\ref{e13}). In the multipole limit, the aspect ratio $q$ 
of the nanoshell plays the role of the coupling constant between the solid and cavity sphere plasmons. Hence, by 
setting $q = 0$ in the symmetric mode $(\omega_{l-})$ of Eq. (\ref{e13}), one finds the multipole LSPR of a solid sphere:
\begin{eqnarray}
\omega^{s}_{l} = \omega_{l-}(q = 0)~~~~~~~~~~~~~~~~~~~~~\nonumber \\
= \omega_{p}\sqrt{\frac{l}{l\varepsilon_{\infty} + (l+1)\varepsilon_{m}}}\label{e21}.~~~~
\end{eqnarray}
The dipole limit of Eq. (\ref{e21}) can be obtained by simply setting $l=1$, or by performing 
\textbf{S} of Eq. (\ref{e15}), where the core-volume fraction $f_{c}$, which plays the role of the 
coupling constant between the solid and cavity sphere plasmons, has to be set to zero. 
Also, \textbf{GR} can be used by substituting Eq. (\ref{e14}) into Eq. (\ref{e18}), to obtain: 
\begin{eqnarray}
\omega^{s}_{1} = \omega_{1-}(f_{c} = 0)~~~~~~~~~~~~~~~~~~~~~\nonumber \\
= \omega_{p}\sqrt{\frac{1}{\varepsilon_{\infty} + 2\varepsilon_{m}}}\label{e22}.~~~~~~~~~~~~~~~~
\end{eqnarray}
Note that the $\varepsilon_{m} = \varepsilon_{\infty} = 1$ limit of  
Eq. (\ref{e22}) can also be obtained by performing \textbf{S} of Eq. (\ref{e16}).

\subsubsection{Cavity Sphere}
We used \textbf{DR} to obtain the LSPR of a cavity sphere. 
To do this, the expression obtained for $\Re[\varepsilon(\omega_{r})]$ in the case of a solid sphere
is used, but the positions of the dielectric constants are reversed, i.e.,
$\varepsilon_{m}$ is replaced with $\varepsilon_{c}$, to yield an expression for 
$\Re[\varepsilon(\omega_{r})]$ (see SII.2, ESI), which is substituted in Eq. (\ref{e2}), providing the following expression for the 
longitudinal multipole LSPR of the cavity sphere:
\begin{eqnarray}
\omega^{c}_{l} = \omega_{l+}(q=0)~~~~~~~~~~~~~~~~\nonumber \\
=  \omega_{p}\sqrt{\frac{l+1}{l\varepsilon_{c} + (l+1)\varepsilon_{\infty}}}.\label{e23}
\end{eqnarray}
However, Eq. (\ref{e23}) can also be obtained by performing \textbf{AS} of Eq. (\ref{e13}), i.e., by setting 
$q = 0$ in the antisymmetric mode $(\omega_{l+})$ of Eq. (\ref{e13}).
The dipole limit of Eq. (\ref{e23}) can be obtained by simply setting $l=1$, or by 
performing \textbf{AS} of Eq. (\ref{e15}), or 
by using \textbf{GR}, i.e., by substituting Eq. (\ref{e14}) into Eq. (\ref{e20}), giving:
\begin{eqnarray}
\omega^{c}_{1} = \omega_{1+}(f_{c}=0)~~~~~~~~~~~~~~~~\nonumber \\
=  \omega_{p}\sqrt{\frac{2}{\varepsilon_{c} + 2\varepsilon_{\infty}}}.\label{e24}~~~~~~~~~~~~
\end{eqnarray}

\begin{figure}[h!]
\centering 
\includegraphics[width = .51\linewidth]{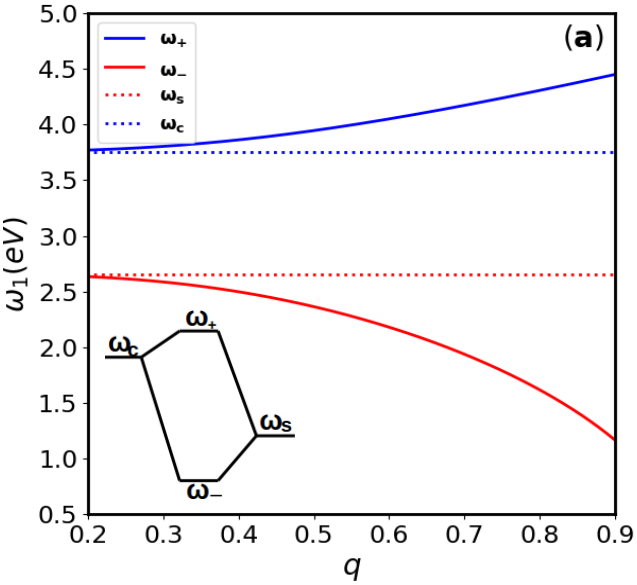}~
\includegraphics[width = .5\linewidth]{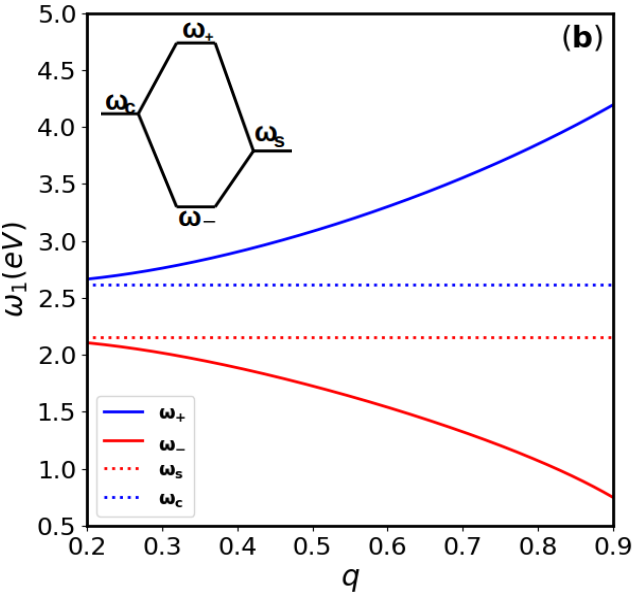}
\caption{\small{Plots of the dipolar symmetric $(\omega_{-})$ and antisymmetric $(\omega_{+})$
modes in a nanoshell against the core aspect ratio $q$, compared with the behaviour of dipolar LSPR of solid and cavity 
sphere plasmons, $\omega_{s}$ and $\omega_{c}$, respectively. 
\textbf{(a)} In the limit $\varepsilon_{m} = \varepsilon_{c} = 1$. 
\textbf{(b)} In the presence of dielectrics $\varepsilon_{m} = 1.78$ (water), $\varepsilon_{c} = 4.2$ (Zirconia).
Insets represent the energy level diagrams for a nanoshell with a core aspect ratio of $0.7$. 
}\label{f3}
}
\end{figure}

\begin{figure}[h!]
\centering 
\includegraphics[width = .5\linewidth]{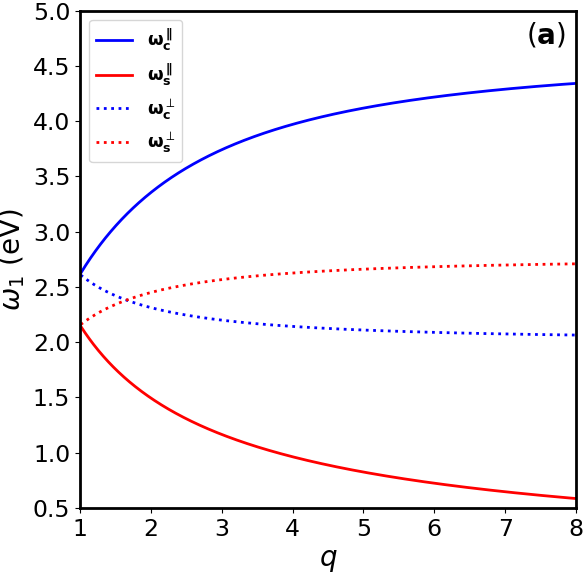}~
\includegraphics[width = .51\linewidth]{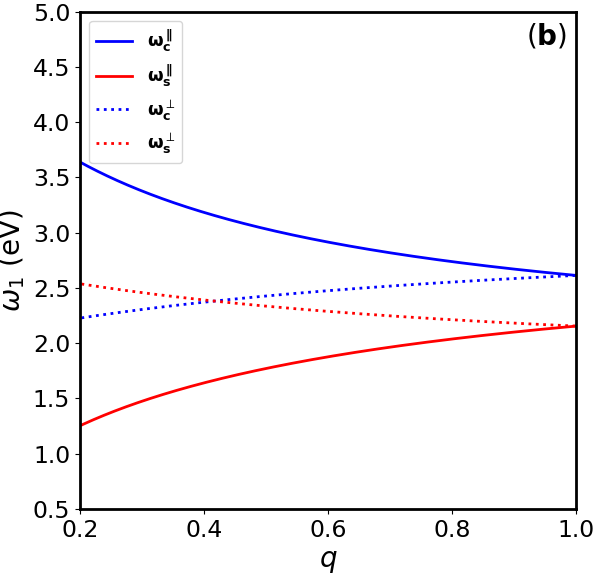}
\caption{\small{The longitudinal and transverse dipolar LSPR of solid and cavity \textbf{(a)} prolate spheroids, and 
\textbf{(b)} oblate spheroids against their aspect ratios, using the following dielectric constants: 
$\varepsilon_{m} = 1.78$, $\varepsilon_{c} = 4.2$. 
}\label{f4}
}
\end{figure}

\begin{figure}[h!]
\centering
\includegraphics[width = .5\linewidth]{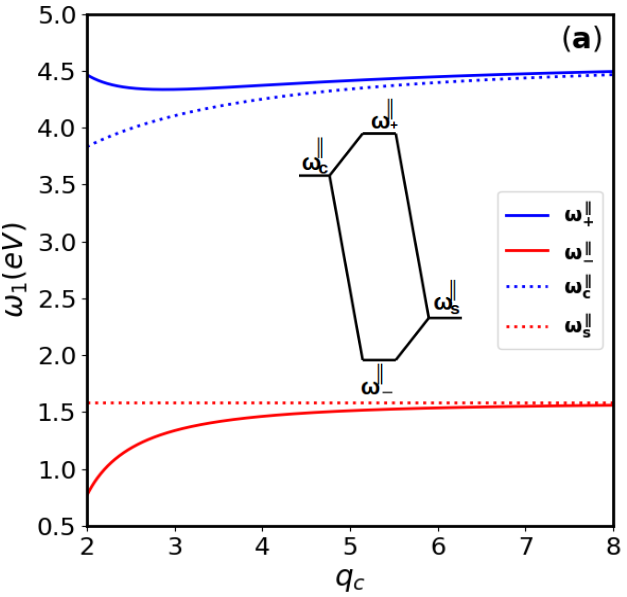}~
\includegraphics[width = .52\linewidth]{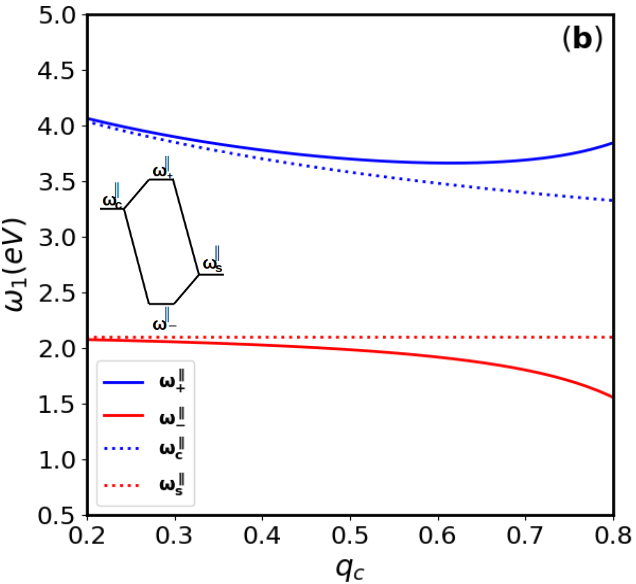}
\caption{\small{Plots of the longitudinal dipolar symmetric $(\omega^{\parallel}_{-})$ and 
antisymmetric $(\omega^{\parallel}_{+})$
modes in \textbf{(a)} a nanorice of aspect ratio $q_{s} = 1.8$,
and \textbf{(b)} a core-shell oblate spheroid of aspect ratio $q_{s} = 0.9$, 
against their different core aspect ratios $q_{c}$, compared with the behaviour of the longitudinal dipolar LSPR of 
the solid and cavity spheroid plasmons, $\omega^{\parallel}_{s}$ and $\omega^{\parallel}_{c}$, respectively. The following 
dielectric constants were used: $\varepsilon_{m} = 1.78$ (water) and $\varepsilon_{c} = 2.1$ (silica). 
Insets represent the energy level diagrams for \textbf{(a)} the nanorice with a core aspect ratio of $q_{c} = 2.0$, and 
\textbf{(b)} the core-shell oblate spheroid with a core aspect ratio of $q_{c} = 0.8$. 
}\label{f5}
}
\end{figure}

FIG. \ref{f3}(a) and FIG. \ref{f3}(b) indicate that an increase in the core aspect ratio $q$ of the nanoshell, i.e., 
thinning of the shells, causes a redshift in the symmetric mode and a blueshift in the antisymmetric mode, while decreasing 
the core aspect ratio, i.e., thickening of the shells, causes the hybrid modes to approach the LSPR of the solid and cavity 
sphere plasmons, respectively. The presence of the dielectric causes a redshift in the LSPR of the nanostructures. 
The energy level diagrams indicate the different energy 
shifts in the LSPR of a nanoshell with a core aspect ratio of $0.7$ (thin shell). The energy shifts decrease with a decrease 
in the core aspect ratio. 

In FIG. \ref{f4}(a), increasing the aspect ratio of a prolate spheroid causes a redshift in both the longitudinal 
LSPR of the solid spheroid and the transverse LSPR of the cavity spheroid. Conversely, the transverse LSPR of the solid 
spheroid and the longitudinal LSPR of the cavity spheroid are both blueshifted. FIG. \ref{f4}(b) shows that this trend is 
reversed in an oblate spheroid, i.e., increasing the aspect ratio of the spheroid affects the longitudinal LSPR of the solid 
spheroid and the transverse LSPR of the cavity
spheroid in the same manner, (i.e., causing a blueshift), while the transverse LSPR of the solid spheroid and
the longitudinal LSPR of the cavity spheroid are both redshifted. However, the LSPR of the solid  
and cavity spheroids approach the LSPR of the solid and cavity sphere, respectively, as $q\longrightarrow 1$. 
A comparison of both plots shows that the longitudinal LSPR of the solid oblate spheroid is 
blueshifted from that of the solid prolate spheroid, while the transverse LSPR of the solid oblate spheroid
is redshifted from that of the solid prolate spheroid, as their aspect ratios increase. The LSPR of the
cavity spheroids shows a converse trend. In addition, both plots show that the longitudinal LSPR of the spheroids 
is more sensitive to an increase in aspect ratio compared to their transverse counterparts. 

FIG. \ref{f5}(a) shows that as the aspect ratio of
the core increases, the symmetric and antisymmetric modes of the nanorice in the longitudinal polarization 
approach the LSPR of the solid and cavity prolate spheroid plasmons, respectively. The transverse 
modes show a similar trend (see S 2, ESI), but the transverse symmetric mode is blueshifted from the 
longitudinal symmetric mode while the transverse antisymmetric mode is redshifted from the longitudinal antisymmetric mode. 
This is because increasing $q_{c}$ at a constant $q_{s}$ causes the core-volume fraction to decrease, i.e., the shells 
become thicker, and the nanorice plasmons will therefore de-couple into the fundamental plasmons.  
In a core-shell oblate spheroid, this trend is reversed, as shown in FIG. \ref{f5}(b), i.e., as the aspect ratio of the 
core decreases, 
the symmetric and antisymmetric modes approach the LSPR of the solid and cavity spheroid plasmons, respectively. 
The energy shifts in the energy level diagrams indicate that the core-shell spheroids with the parameters given in the 
caption consist of thin shells, which increases the coupling strength between the solid and cavity spheroid plasmons.  
Observe that the symmetric modes of the core-shell oblate spheroids are redshifted from those of the nanorice, 
while the antisymmetric modes are blueshifted from those of the nanorice.

\subsection{Simplified Model of Plasmon Hybridization}
In this section, we constructed a simplified model of plasmon hybridization in the core-shell nanostructures we have studied.  
In the core-shell nanostructures, solid plasmons couple to cavity plasmons to form hybrid plasmons, through the core volume fraction $f_{c}$, which plays the role of the coupling constant in the dipole limit. 
This phenomenon is known as plasmon hybridization \cite{Prodan04,Wu06,Wang06}. We used this model to justify the  
de-coupling of the hybrid plasmon modes through $f_{c} = 0$, to form the fundamental plasmon modes, 
as a part of the \textbf{S} and \textbf{AS} methods we have employed in Section \ref{s3.1}. 
For brevity, we only considered the $\varepsilon_{m} = \varepsilon_{c} = \varepsilon_{\infty} = 1$ limit. 

\subsubsection{Core-Shell Sphere}
Consider a solid sphere and a cavity sphere with dimensions as given in Section \ref{s3.1}, but with $b>a$.
When these two nanostructures couple to form a nanoshell, the core-volume fraction of the nanoshell is $f_{c} = (a/b)^{3}$. 
Eq. (\ref{e16}), which gives the normal mode frequencies of plasma oscillations in the nanoshell, can be rewritten as
\begin{equation}\label{e25}
\omega^{2}_{\pm} = \frac{1}{2}\left[(\omega_{c}^{2}+\omega_{s}^{2})
\pm(\omega_{c}^{2}-\omega_{s}^{2})\sqrt{1+8f_{c}} \right],
\end{equation}
where $\omega_{c}$ and $\omega_{s}$ are the dipolar LSPR of the cavity and solid sphere, respectively.
We can construct a simple model for plasmon hybridization by assuming 
that the cavity plasmons couple weakly with the solid plasmons in the nanoshell. 
This weak-coupling regime corresponds to $f_{c}<\frac{1}{8}$, i.e., thick shells, where $f_{c}$ is the coupling constant.
Hence, the weak-coupling regime of solid and cavity plasmons in a nanoshell has an upper bound that corresponds 
to the core-volume fraction $f_{c} = (a/b)^{3} = 1/8$ 
in the limit of $\varepsilon_{m} = \varepsilon_{c} = \varepsilon_{\infty} = 1$. 
From Eq. (\ref{e25}), the normal mode frequencies in the weak-coupling regime up to second order can be obtained: 
\begin{equation}\label{e26}
\omega^{2}_{+} \approx \omega_{c}^{2} + 2f_{c}\left(\omega^{2}_{s}-f_{c}\omega^{2}_{c}\right)
\end{equation}
and 
\begin{equation}\label{e27}
\omega^{2}_{-} \approx \omega_{s}^{2} - 2f_{c}\left(\omega^{2}_{s}-f_{c}\omega^{2}_{c}\right).
\end{equation}

\begin{figure}[h!]
\centering 
\includegraphics[width = .6\linewidth]{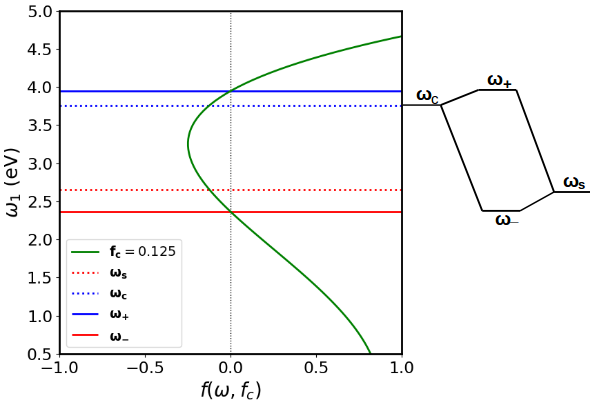}
\caption{\small{ Graphical solution of the characteristic function 
$f(\omega,f_{c}) = \omega^{4}-\omega^{2}(\omega^{2}_{+}+\omega^{2}_{-})+\omega^{2}_{+}\omega^{2}_{-}$, 
at the upper bound $f_{c} = 0.125$ of the weak-coupling regime  ($f_{c} < 0.125$). 
The zeros of $f(\omega,0.125)$ are at $\omega_{-} = 2.35$~eV (red solid line) and $ \omega_{+} = 3.95$~eV (blue solid line), 
at the LSPR $f(\omega = \omega_{r},f_{c}) = 0$. The function has a minimum value at $\omega = \omega_{p}/\sqrt{2}$.
Eqs. (\ref{e26}) and (\ref{e27}) predict $\omega_{-} = 2.39$~eV and $ \omega_{+} = 3.92$~eV, respectively. 
The zeros of $f(\omega,f_{c}< 0.125)$ will approach $\omega_{s}$ and $\omega_{c}$, respectively.  
The energy shifts are illustrated in the energy level diagram on the right.  
}
}\label{f6}
\end{figure}

\subsubsection{Core-Shell Spheroid}
Consider a solid prolate spheroid and a cavity prolate spheroid with dimensions as given in Section \ref{s3.1}, 
but with $q_{c}>q_{s}$. When these two nanostructures couple to form a nanorice, the core-volume fraction of the 
nanorice is $f_{c}$ as given in Section \ref{s3.1.1}. Eq. (\ref{e11}), giving the following normal modes of the plasma 
oscillations in the nanorice for the longitudinal polarization:
\begin{equation}\label{e28}
\resizebox{.6\hsize}{!}{$
\omega^{2}_{\pm} = \frac{1}{2}
\left[(\omega_{c}^{2}+\omega_{s}^{2})\pm (\omega_{c}^{2}-\omega_{s}^{2})
\sqrt{ 1 + \frac{ 4f_{c}L_{s}^{\parallel}\Big(1-L_{s}^{\parallel}\Big) }{ \Big(1-L_{c}^{\parallel}-L_{s}^{\parallel}\Big)^{2} }}  
~\right],
$}
\end{equation}
where $\omega_{c}$ and $\omega_{s}$ are the longitudinal dipolar LSPR of the cavity and solid prolate
spheroid, respectively, for which the symbol ``$\parallel$'' has been dropped. 

We can construct a simple model of plasmon hybridization by assuming that the cavity plasmons couple weakly with
the solid plasmons in the nanorice. This weak-coupling regime corresponds to 
$f_{c} < \Big[\Big(1-L_{c}^{\parallel}-L_{s}^{\parallel}\Big)^{2}/4L_{s}^{\parallel}\Big(1-L_{s}^{\parallel}\Big)\Big]$. 
Hence, the weak-coupling regime of solid and cavity plasmons in a nanorice has an upper bound that corresponds 
to the core-volume fraction $f_{c} = v_{c}(v^{2}_{c}-1)/v_{s}(v^{2}_{s}-1) = 
(1-L_{c}^{\parallel}-L_{s}^{\parallel})^{2}/4L_{s}^{\parallel}(1-L_{s}^{\parallel})$  
in the limit of $\varepsilon_{m} = \varepsilon_{c} = \varepsilon_{\infty} = 1$.

From Eq. (\ref{e28}), the normal mode frequencies in the weak-coupling regime up to the second order can be obtained, giving: 
\begin{equation}\label{e29}
\resizebox{.6\hsize}{!}{$
\omega^{2}_{+} \approx 
\omega_{c}^{2}+f_{c}\left(\frac{1-L_{s}^{\parallel}}{1-L_{c}^{\parallel}-L_{s}^{\parallel}}\right)
\left[\omega^{2}_{s}-f_{c}\omega^{2}_{c}
\left(\frac{L_{s}^{\parallel}}
{1-L_{c}^{\parallel}-L_{s}^{\parallel}}
\sqrt{ \frac{ 1-L^{\parallel}_{s} }{ 1-L^{\parallel}_{c} } }~\right)^{2}~\right]
$}
\end{equation}
and
\begin{equation}\label{e30}
\resizebox{.6\hsize}{!}{$
\omega^{2}_{-} \approx 
\omega_{s}^{2}-f_{c}\left(\frac{1-L_{s}^{\parallel}}{1-L_{c}^{\parallel}-L_{s}^{\parallel}}\right)
\left[\omega^{2}_{s}-f_{c}\omega^{2}_{c}
\left(\frac{L_{s}^{\parallel}}
{1-L_{c}^{\parallel}-L_{s}^{\parallel}}
\sqrt{ \frac{ 1-L^{\parallel}_{s} }{ 1-L^{\parallel}_{c} } }~\right)^{2}~\right].
$}
\end{equation}
\begin{figure}[h!]
\centering 
\includegraphics[width = .6\linewidth]{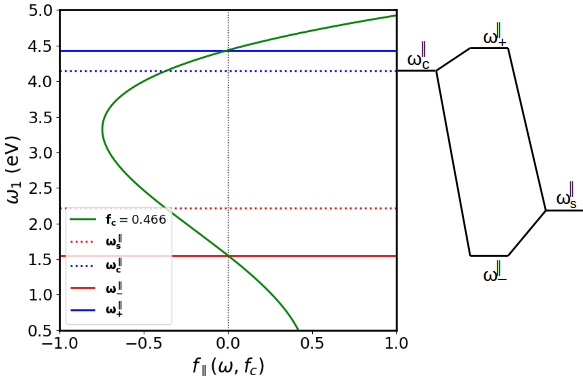}
\caption{\small{Graphical solution of the characteristic function
$f_{\parallel}(\omega,f_{c}) = \omega^{4}-\omega^{2}(\omega^{2}_{+}+\omega^{2}_{-})+\omega^{2}_{+}\omega^{2}_{-}$, 
for a nanorice of aspect ratio $q_{s} = 1.5$, and core aspect ratio $q_{c} = 1.839$, which corresponds to an 
upper bound of $f_{c} = 0.466$ in the weak-coupling regime ($f_{c}<0.466$). 
The zeros of $f_{\parallel}(\omega,0.466)$ are $\omega^{\parallel}_{-} = 1.53$~eV and $ \omega^{\parallel}_{+} = 4.43$~eV 
at the LSPR $f_{\parallel}(\omega = \omega_{r},f_{c}) = 0$ for the longitudinal polarization. 
The function has a minimum at $\omega = \frac{\omega_{p}}{\sqrt{2}}\sqrt{1-L_{c}^{\parallel}+L_{s}^{\parallel}}$. 
Eqs. (\ref{e29}) and  Eq. (\ref{e30}) predict $\omega^{\parallel}_{-} = 1.62$~eV and $ \omega^{\parallel}_{+} = 4.41$~eV, respectively. 
The zeros of $f_{\parallel}(\omega,f_{c}< 0.466)$ will therefore approach $\omega^{\parallel}_{s}$ and 
$\omega^{\parallel}_{c}$, respectively. The energy shifts are shown in an energy level diagram on the right. 
}
}\label{f7}
\end{figure}
\\
From the above simple models of plasmon hybridization in a nanoshell and in a nanorice, 
we can make the following deductions: 
Eqs. (\ref{e26}) and (\ref{e29}) and Eqs. (\ref{e27}) and (\ref{e30}) decouple into the dipolar LSPR of cavity 
and solid plasmons, respectively, once $f_{c}$ is set to zero,  i.e., once we perform \textbf{AS} and \textbf{S}, respectively. 
The weak-coupling regime in nanoshells has a unique upper bound of 
$\frac{1}{8}$ in the limit $\varepsilon_{\infty} = \varepsilon_{m} = \varepsilon_{c} = 1$, 
so that different nanoshells have the same weak-coupling regime, irrespective of their differing aspect ratios. 
The weak-coupling regime in nanorice does not have a unique upper bound, since it depends on the geometric factors 
$L_{s}^{\parallel}$ and $L_{c}^{\parallel}$. This means that different nanorice, will have different weak-coupling regimes. 
However, in the presence of dielectrics the weak-coupling regime in a nanoshell will no longer have a unique 
upper bound because it will depend on the choice of dielectrics used. The energy level diagrams on the right 
of FIG. \ref{f6} and FIG. \ref{f7} respectively, show the energy shifts at the upper bounds of the weak-coupling regime.
These energy shifts will further decrease below the upper bounds, causing the symmetric and antisymmetric modes to 
approach the LSPR of the solid and cavity plasmons respectively. 
Finally, in the weak-coupling regime, the contribution from the solid plasmons dominates the normal mode frequencies 
of the hybrid system. This is expected since the weak-coupling regime is also the thick-shell limit.

\subsection{Beyond the Drude Model and the Rayleigh Regime}
The formulas we have presented in the above sections are easily reproducible.  
Though, we have considered gold nanostructures only, the formulas 
can also be used to predict the LSPR of other nanostructures of Drude metals, such as
silver nanostructures, once the right parameters have been chosen.  

Nevertheless, corrections due to retardation effects \cite{Moroz09,Ford84}, 
non-local effects \cite{Raza14,Kreig85}, size-dependent response \cite{Moroz09,Raza14,Anasta16}, 
interband effects \cite{Abajo08,Anasta16}, and surface roughness \cite{Martin12,Ford84} have to be carried out 
on these formulas for accurate prediction of the LSPR. 
Also, higher order LSPR, such as the quadrupolar LSPR (due to $l = 2$ plasmons) becomes non-negligible beyond the Rayleigh regime, 
and has to be evaluated \cite{Anasta16,Dmitri13}.

Interband transitions in gold can be accounted for through an extra imaginary part added to 
Eq. (\ref{e1}) \cite{Anasta16}, which is a good alternative to the use of multiple Lorentzian functions. Retardation effects 
such as dynamic polarization and radiation damping are usually accounted for by adding correction terms to the 
static geometric factors \cite{Moroz09,Ford84}. Size-dependent damping of the Kreibig type \cite{Raza14,Kreig85} 
is usually accounted for in solid spherical nanoparticles by adding a size-dependent damping term to the 
free-electron damping rate in Eq. (\ref{e1}). Surface roughness can be accounted for through a roughness gain \cite{Ford84}
that depends on the roughness features. Non-local effects such as electron convention and diffusion
are accounted for through a wavevector-dependent complex dielectric response of the metal \cite{Raza14}. 
In gold nanostructures, the predominant effect of these corrections on the LSPR calculated using the LRA is a redshift. 
However, the overall process of carrying out these corrections on the formulas derived here will involve a less distinct 
mathematical formalism.

\section{Conclusions}
We have identified and discussed some important relationships between simple plasmonic nanostructures.
We believe that these relationships extend to other plasmonic nanostructures, 
especially to nanostructures where hybrid plasmons are present. 
However, we have limited our discussion to
nanostructures with smooth corners in the Rayleigh regime, where both an analytical approach and a
quasistatic description are possible. We derived their dipolar and multipolar LSPR 
using a combination of the Drude model, the Rayleigh approximation, and the 
Fr\"{o}hlich condition.
Our results have shown that the LSPR depends strongly on the geometric factors of the nanostructures,
which emphasizes the shape dependence of the LSPR. 
The approach we adopted led us to several formulas for 
calculating the LSPR of the most common nanostructures in plasmonics literature. 
Some of these formulas have not been reported previously, to the best of our knowledge, especially in the 
manner in which we have simplified them using geometric factors and the mutual relationships between nanostructures. 
Beside LSPR, these relationships will most likely exist in other optical properties of the nanostructures.
The formulas we have derived are handy, and can easily be used for estimating the LSPR of single metallic 
nanostructures in different environments.

\begin{acknowledgments}
L. C. U. was sponsored by the National Research Foundation (NRF) and the University of Pretoria.
T. M. was supported by the Czech Science Foundation (GACR) grant no. 17-22160S. 
T. P. J. K. was supported by the NRF project nos. 109302 and 112085.
We wish to thank Vincenzo Giannini, for his advice on the manuscript. 
\end{acknowledgments}

\bibliography{Manuscript}

\end{document}